\documentstyle[12pt]{article}
\begin{document}

\title{Energy and Angular Momentum Densities in a G\"{o}del-Type Universe in
the Teleparallel Geometry}
\author{A. A. Sousa*, R. B. Pereira and A. C. Silva \\
Instituto de Ci\^{e}ncias Exatas e da Terra\\
Campus Universit\'{a}rio do Araguaia\\
Universidade Federal de Mato Grosso\\
78698-000 Pontal do Araguaia, MT, Brazil\\
}
\maketitle

\begin{abstract}
The main scope of this research consists in evaluating the energy-momentum
(gravitational field plus matter) and gravitational angular momentum
densities in the universe with global rotation, considering the G\"{o}%
del-Obukhov metric. For this, we use the Hamiltonian formalism of the
Teleparallel Equivalent of General Relativity (TEGR), which is justified for
presenting covariant expressions for the considered quantities. We found
that the total energy density calculated by the TEGR method is in agreement
with the results reported by other authors in the literature using
pseudotensors. The result found for the angular momentum density depends on
the rotational parameter as expected. We also show explicitly the
equivalence among the field equations of the TEGR and Einstein equations
(RG), considering a perfect fluid and G\"{o}del-Obukhov metric.

PACS NUMBERS: 98.80.-k, 04.20.-q, 04.20.Cv, 04.20.Fy

(*) E-mail: adellane@ufmt.br
\end{abstract}

\section{Introduction}

Determination of the energy, momentum, and angular momentum of the
gravitational field is a long-standing problem in General Relativity (GR).
In a general way, it is believed that the energy of the gravitational field
is not localizable, that is, defined in a finite region of space. This
problem is addressed in the literature by means of different approaches. The
most common approach is based in the use of pseudotensors, since these
gravitational physical quantities do not possess proper definitions in terms
of tensorial equations [1].

In order to consider the material content of space-time, it is possible to
define the energy-momentum (angular momentum) complex as the sum of the
energy-momentum (angular momentum) pseudotensor of the gravitational field
and the energy-momentum (angular momentum) tensor of the matter. These
complexes appear with several names, such as Landau-Lifshitz,
Bergman-Thompson, and Einstein, among others [2]. They differ from each
other in the way in which they are constructed. These complexes have been
applied to several configurations of the gravitational field, as
Friedmann-Lema\^{\i}tre-Robertson-Walker (FLRW) [3]--[7], Bianchi I and II
[8], [9], G\"{o}del [10], and G\"{o}del-type universes [11] and for a
general non-static spherically symmetric metric of the Kerr-Schild class
[12].

In an attempt to contour the localized gravitational energy problem, we
consider the Weitzenb\"{o}ck space-time, a particular case of Riemann-Cartan
space-time constrained to have zero curvature. In this space-time, the
gravitational field is described in terms of the tetrad field $e^{a}{}_{\mu
} $, instead of metric tensor $g_{\mu \nu }.$ In the context of teleparallel
gravity, the zero curvature condition means that the parallel transport of
the orientation of the tetrads is path independent, that is, there is a
global moving frame or teleparallelism. Mathematically, as a natural
consequence of this definition, the Weitzenb\"{o}ck covariant derivative of
the tetrad field vanishes identically: $\bigtriangledown _{\nu }e_{a\mu
}\equiv \partial _{\nu }e_{a\mu }-\Gamma _{\nu \mu }^{\rho }e_{a\rho }=0,$
where $\Gamma _{\nu \mu }^{\rho }$ is the Weitzenb\"{o}ck connection given
by $\Gamma _{\nu \mu }^{\rho }=e_{a}^{\;\rho }\partial _{\nu }e^{a}{}_{\mu
}. $ The gravitational interaction is attributed to the torsion tensor $%
T_{a\mu \nu }${\bf = }$\partial _{\mu }e_{a\nu }${\bf \ }$-${\bf \ }$%
\partial _{\nu }e_{a\mu }.$ In Andrade \& Pereira [13], the teleparallel
gravity can indeed be understood as a gauge theory for the translation
group. In that approach, the gravitational interaction is described by a
force similar to the Lorentz force equation of electrodynamics, with torsion
playing the role of force.

In 1961, M\o ller [14] showed that only in terms of tetrads can we obtain a
Lagrangian density that leads to a tensor of gravitational energy-momentum.
This tensor, constructed from the first derivatives of the tetrads, does not
annul any coordinate's transformation.

In 2001, Maluf established the Hamiltonian formulation of the Teleparallel
Equivalent of General Relativity (TEGR) without the Schwinger time gauge. In
this formulation, the definition of the energy and gravitational angular
momentum arises by suitably interpreting the integral forms of the
constraints' equations $C^{a}=0$ and $\Gamma ^{ab}=0,$ respectively [15],
[16]. Several configurations of gravitational energies were investigated
with success, such as the space-time configurations of de Sitter [17],
conical defects [18], static Bondi [19], disclination defects [20], Kerr
black hole (irreducible mass) [21], BTZ black hole [22], and Kerr anti-de
Sitter [23]. The definition of gravitational angular momentum as the
integral form of the constraint equation was applied satisfactorily for the
gravitational field of a thin, slowly rotating mass shell [16] and for the
rotational BTZ black hole [24].

On the other hand, the idea that the Universe is rotating was first
mentioned by Gamow [25]. The exact solution of the Einstein equations for
the model of a homogeneous, static universe with rotation was proposed for G%
\"{o}del [26]. However, the G\"{o}del equations presented serious problems,
such as closed time-like curves (CTCs) that exhibit a causality violation in
space-time. Several new models based on G\"{o}del's original idea were
created in an attempt to address the problems in their model [27]--[29].
Among these ideas, we stress the nonstationary model proposed by Obukhov
[30], which presents expansion as well as rotation. This generalization is
known as the G\"{o}del-Obukhov or G\"{o}del-type anisotropic cosmological
model [31]. This model would not be in conflict with any cosmological
observations [32].

Despite advances in the theoretical understanding of these problems, the
global rotation has not yet been detected. However, we should stress that
there are weak observational evidences of global rotation of the universe
[33].

In this work, we show explicitly the equivalence between GR and TEGR. More
specifically, we consider the solution for an anisotropic and homogeneous
universe described by the G\"{o}del-Obukhov metric in Cartesian coordinates.
The main reason for using these coordinates is to allow posterior comparison
of our work with other literature results. We demonstrate that considering a
perfect fluid, there is no solution that incorporates rotation and expansion
simultaneously that is in accord with Ozsv\'{a}th [34]. By making the
rotational parameters equals to zero, we obtain the Friedmann equations of
FLRW for a spatially flat universe.

We also apply the Hamiltonian formulation and field equations of the TEGR
implemented by Maluf [15], [16] to find the energy-momentum (gravitational
field plus matter) and gravitational angular momentum densities in the G\"{o}%
del-Obukhov universe, irrespective of the equations of the state of the
cosmic fluid. Our result for total energy density is in accord with that
obtained by Rybn\'{\i}ckov\'{a}, differing by a constant factor, where the
Komar superpotential was used [11]. Our result for gravitational angular
momentum disagree with that obtained by Dabrowski \& Garecki using the
gravitational angular momentum pseudotensor of Bergman-Thomson [10].

The article is organized as follows. In Section 2 we review the Lagrangian
and Hamiltonian formulation of the TEGR. In Section 3, using the field
equations of the TEGR, we find the teleparallel version of the Friedmann
equations. In Section 4, we calculate the total energy of the G\"{o}%
del-Obukhov universe and compare it with those obtained from the
pseudotensors. In Section 5, we find the total three-momentum of the
universe. In Section 6, we also obtain the gravitational angular momentum
density of this G\"{o}del-type universe. Finally, in Section 7, we present
our conclusions.

The notation is the following: space-time indices $\mu $,$\nu $, ... and
global SO(3, 1) indices $a,b$,... run from 0 to 3. Time and space indices
are indicated by $%
{\mu}%
=0,a=(0),(i)$. The flat, Minkowski space-time metric tensor raises and
lowers tetrad indices and is fixed by $\eta _{ab}=e_{a\mu }e_{b\nu }g^{\mu
\nu }=(-+++)$. The determinant of the tetrad field is represented by $%
e=det(e_{\quad \mu }^{a})$. We use units in which $c=1$, where $c$\ is the
light speed.

\section{The Hamiltonian constraints equations as an energy and
gravitational angular momentum equations}

We will briefly recall both the Lagrangian and Hamiltonian formulations of
the TEGR. The Lagrangian density for the gravitational field in the TEGR
[35] with the cosmological constant $\Lambda $\ is given by

\begin{eqnarray}
L(e_{a\mu }) &=&-k^{\prime }\,e\,({\frac{1}{4}}T^{abc}T_{abc}+{\frac{1}{2}}%
T^{abc}T_{bac}-T^{a}T_{a})\;-L_{M}-2ek^{\prime }\Lambda  \nonumber \\
&\equiv &-k^{\prime }e\Sigma ^{abc}T_{abc}-L_{M}-2ek^{\prime }\Lambda ,
\end{eqnarray}%
where $k^{\prime }$\ $=1/(16\pi G$), $G$\ is the Newtonian gravitational
constant, and stands for the Lagrangian density for the matter fields. As
usual, tetrad fields convert space-time into Lorentz indices and vice-versa.
The tensor $\Sigma ^{abc}$\ is defined by%
\begin{equation}
\Sigma ^{abc}={\frac{1}{4}}(T^{abc}+T^{bac}-T^{cab})+{\frac{1}{2}}(\eta
^{ac}T^{b}-\eta ^{ab}T^{c}),
\end{equation}%
and $T^{b}=T^{a}\,_{a}\,^{b}.$The quadratic combination $\Sigma
^{abc}T_{abc} $\ is proportional to the scalar curvature $R(e)$, except for
a total divergence. The field equations for the tetrad field read

\begin{equation}
e_{a\lambda }e_{b\mu }\partial _{\nu }(e\Sigma ^{b\lambda \nu })-e\biggl(%
\Sigma ^{b\nu }\,_{a}T_{b\nu \mu }-{\frac{1}{4}}e_{a\mu }T_{bcd}\Sigma ^{bcd}%
\biggr)+\frac{1}{2}ee_{a\mu }\Lambda =\frac{1}{4k^{\prime }}eT_{a\mu }\;.
\label{5000}
\end{equation}%
where $eT_{a\mu }=\delta L_{M}/\delta e^{a\mu }$. It is possible to prove by
explicit calculations that the left hand side of Eq. (3) is exactly given by

\begin{equation}
{\frac{1}{2}}\,e\,\biggl\{R_{a\mu }(e)-{\frac{1}{2}}e_{a\mu }R(e)+e_{a\mu
}\Lambda \biggr\}\;,  \label{4000}
\end{equation}%
aand thus it follows that the field equations arising from the variation in $%
L$\ with respect to $e^{a}\;_{\mu }$\ are strictly equivalent to Einstein's
equations in tetrad form.

The field equations (3) may be rewritten in the form

\begin{equation}
\partial _{\nu }\left( e\Sigma ^{a\lambda \nu }\right) =\frac{1}{4k^{\prime }%
}ee^{a}{}_{\mu }\left( t^{\lambda \mu }+\tilde{T}^{\lambda \mu }\right) \;,
\label{16000}
\end{equation}%
where

\begin{equation}
t^{\lambda \mu }=k^{\prime }\left( 4\Sigma ^{bc\lambda }T_{bc}{}^{\mu
}-g^{\lambda \mu }\Sigma ^{bcd}T_{bcd}\right) \;,
\end{equation}%
and

\begin{equation}
\tilde{T}^{\lambda \mu }=T^{\lambda \mu }+2k^{\prime }g^{\lambda \mu
}\Lambda ,
\end{equation}%
are interpreted as the gravitational energy-momentum tensor [36] and the
matter energy-momentum tensor, respectively.

The Hamiltonian formulation of the TEGR is obtained by first establishing
the phase space variables. The Lagrangian density does not contain the time
derivative of the tetrad component . Therefore this quantity will arise as a
Lagrange multiplier [37]. The momentum canonically conjugated to $e_{ai}$\
is given by $\Pi ^{ai}=\delta L/\delta \dot{e}_{ai}.$\ The Hamiltonian
formulation is obtained by rewriting the Lagrangian density in the form $L=p%
\dot{q}-H$\ , in terms of $e_{ai}$, $\Pi ^{ai}$\ and Lagrange multipliers.
The Legendre transform can be successfully carried out, and the final form
of the Hamiltonian density is expressed as [15]%
\begin{equation}
H=e_{a0}C^{a}+\alpha _{ik}\Gamma ^{ik}+\beta _{k}\Gamma ^{k},  \label{15000}
\end{equation}%
plus a surface term. Here $\alpha _{ik}$\ and $\beta _{k}$\ are Lagrange
multipliers that (after solving the field equations) are identified as $%
\alpha _{ik}$\ =1/2($T_{i0k}$\ + $T_{k0i}$) and $\beta _{k}$\ = $T_{00k}$\ $%
. $\ $C^{a},$\ $\Gamma ^{ik}$\ and $\Gamma ^{k}$\ are first class
constraints. The Poisson bracket between any two field quantities F and G is
given by

\begin{equation}
\{F,G\}=\int d^{3}x\biggl(\frac{{\delta F}}{{\delta e_{ai}(x)}}\frac{{\delta
G}}{{\delta \Pi ^{ai}(x)}}-\frac{{\delta F}}{{\delta \Pi ^{ai}(x)}}\frac{{%
\delta G}}{{\delta e_{ai}(x)}}\biggr)\;.
\end{equation}

The constraint $C^{a}$\ is written as $C^{a}=-\partial _{i}\Pi ^{ai}+h^{a}$,
where $h^{a}$\ is an intricate expression of the field variables. The
integral form of the constraint equation $C^{a}$\ $=0$\ motivates the
definition of the energy-momentum four-vector [15]

\begin{equation}
P^{a}=-\int_{V}d^{3}x\partial _{i}\Pi ^{ai}\;,
\end{equation}%
where $V$\ is an arbitrary volume of the three-dimensional space. In the
configuration space we have

\begin{equation}
\Pi ^{ai}=-4k^{\prime }e\Sigma ^{a0i}\;.
\end{equation}

The emergence of total divergences in the form of scalar or vector densities
is possible in the framework of theories constructed out of the torsion
tensor. Metric theories of gravity do not share this feature.

By making $\lambda =0$\ in Eq. (5) and identifying $\Pi ^{ai}$\ on the left
hand side of the latter, the integral form of Eq. (5) can be written as

\begin{equation}
P^{a}=\int_{V}d^{3}xee^{a}{}_{\mu }\left( t^{0\mu }+\tilde{T}^{0\mu }\right)
.  \label{1000}
\end{equation}%
This equation suggests that $P^{a}$\ is now understood as the total,
gravitational, and matter fields (plus a cosmological constant fluid)
energy-momentum [36]. The spatial components $P^{(i)}$\ form a total
three-momentum, while a temporal component $P^{(0)}$\ is the total energy
(gravitational field plus matter) [1].

It is important to rewrite the Hamiltonian density $H$\ in the most simple
form. It is possible to simplify the constraints which may be rewritten as a
single constraint $\Gamma ^{ab}.$\ The Hamiltonian density (8) may be
written in the equivalent form [16]

\begin{equation}
H=e_{a0}C^{a}+\frac{1}{2}\lambda _{ab}\Gamma ^{ab},
\end{equation}%
where $\lambda _{ab}=-\lambda _{ba}$\ are Lagrange multipliers that are
identified as $\lambda _{ik}=\alpha _{ik}$\ and $\lambda _{0k}=-\lambda
_{k0}=\beta _{k}.$\ The constraints $\Gamma ^{ab}=-\Gamma ^{ba}$\ embody
both constraints $\Gamma ^{ik}$\ and $\Gamma ^{k}$\ by means of relations

\begin{equation}
\Gamma ^{ik}=e_{a}{}^{i}e_{b}{}^{k}\Gamma ^{ab},
\end{equation}%
and 
\begin{equation}
\Gamma ^{k}\equiv \Gamma ^{0k}=\Gamma ^{ik}=e_{a}{}^{0}e_{b}{}^{k}\Gamma
^{ab}.
\end{equation}%
It reads

\begin{equation}
\Gamma ^{ba}=M^{ab}+4k^{\prime }e\left( \Sigma ^{a0b}-\Sigma ^{b0a}\right) .
\end{equation}

Similarly to the definition of $P^{a}$, the integral form of the constraint
equation $\Gamma ^{ab}=0$\ motivates the new definition of the space-time
angular momentum. The equation $\Gamma ^{ab}=0$\ implies

\begin{equation}
M^{ab}=-4k^{\prime }e\left( \Sigma ^{a0b}-\Sigma ^{b0a}\right) .
\label{13000}
\end{equation}

Therefore Maluf defines [16]

\begin{equation}
L^{ab}=\int_{V}d^{3}xe^{a}{}_{\mu }e^{b}{}_{\nu }M^{\mu \nu },  \label{12000}
\end{equation}%
where

\begin{equation}
M^{ab}=e^{a}{}_{\mu }e^{b}{}_{\nu }M^{\mu \nu }=-M^{ba}.  \label{14000}
\end{equation}%
as the four-angular momentum of the gravitational field.

The quantities $P^{a}$\ and $L^{ab}$\ are separately invariant under general
coordinate transformations of the three-dimensional space and under time
reparametrizations, which is an expected feature since these definitions
arise in the Hamiltonian formulation of the theory. Moreover these
quantities transform covariantly under global SO(3,1) transformations.

\section{Teleparallel version of G\"{o}del-Obukhov equations}

The line element of the G\"{o}del-Obukhov or G\"{o}del-type cosmological
model is given by the interval [30] {\bf \ }%
\begin{equation}
ds^{2}=-dt^{2}+2a(t)\sqrt{\sigma }e^{mx}dtdy+a(t)^{2}\left(
dx^{2}+ke^{2mx}dy^{2}+dz^{2}\right) .  \label{20000}
\end{equation}%
where $m,$ $\sigma $ and $k$ are the constant parameters and $a(t)$ is the
time-dependent cosmological scale factor. Here, we adopted the signature $%
(-+++).$ A general analysis [30], [38] of the kinematic properties of such a
space-time shows that the closed time-like curves are absent in this
manifold when $k>0$ ($\sigma >0$ by definition). We obtain the usual G\"{o}%
del metric [26] by making \ $a(t)=1$, $\sigma =1,$\ $m=1$\ and $k=-1/2$\ in
the expression (20) .

Using the relations 
\begin{equation}
g_{\mu \nu }=e^{a}{}_{\mu }e_{a\nu },  \label{eq36}
\end{equation}%
and

\begin{equation}
e_{a\mu }=\eta _{ab}e^{b}{}_{\mu },
\end{equation}%
a set of tetrads fields that satisfy the metric is given by [39]

\begin{equation}
e^{a}{}_{\mu }=\left( 
\begin{array}{cccc}
1 & 0 & -a(t)\sqrt{\sigma }e^{mx} & 0 \\ 
0 & a(t) & 0 & 0 \\ 
0 & 0 & a(t)\sqrt{k+\sigma }e^{mx} & 0 \\ 
0 & 0 & 0 & a(t)%
\end{array}%
\right) .  \label{30000}
\end{equation}

This set of tetrads fields yields the velocity field given by $e_{\left(
0\right) }{}^{\mu }$ $(t,x,y,z)=\left( 1,0,0,0\right) $. According to the
physical interpretation of Eq. (23), the latter is adapted to static
observers in space-time [16].

Now, with the help of the inverse metric tensor $g^{\mu \nu }$, we can write
the inverse tetrads%
\begin{equation}
e_{a}{}^{\mu }=g^{\mu \nu }e_{a\nu }\;,
\end{equation}%
as 
\begin{equation}
e_{a}{}^{\mu }=\left( 
\begin{array}{cccc}
1 & 0 & 0 & 0 \\ 
0 & 1/a(t) & 0 & 0 \\ 
\sqrt{\frac{\sigma }{k+\sigma }} & 0 & \frac{e^{-mx}}{a(t)\sqrt{k+\sigma }}
& 0 \\ 
0 & 0 & 0 & 1/a(t)%
\end{array}%
\right) .  \label{31000}
\end{equation}%
where the determinant of $e^{a}{}_{\mu }$\ is%
\begin{equation}
e=a(t)^{3}\sqrt{k+\sigma }e^{mx}.  \label{eq313b}
\end{equation}

Before solving the field equations, it is necessary to consider the material
content of the universe. We restrict our consideration here to the
stress-energy-momentum tensor of a perfect fluid [40] given by 
\begin{equation}
T_{\nu \mu }=\left( \rho +p\right) u_{\nu }u_{\mu }+pg_{\nu \mu },
\end{equation}%
where $\rho =\rho (x)$\ is the matter energy density, $p$\ is the matter
pressure and $u_{\mu }$\ are velocity components. Here, we denote the
velocity with respect to the commoving matter by 
\[
u^{\alpha }=\delta _{t}^{\alpha }, 
\]%
which in covariant notation is written as \ $u_{\mu }=g_{\mu \alpha }\delta
_{t}^{\alpha }.$\ Thus, we can write the nonzero components of the
stress-energy-momentum tensor $T_{\mu \nu }$\ as

\begin{eqnarray}
T_{00} &=&(\varepsilon +p)u_{0}u_{0}+pg_{00}=\varepsilon ,  \label{32000} \\
T_{02} &=&(\varepsilon +p)u_{0}u_{2}+pg_{02}=T_{20}=-a\sqrt{\sigma }%
e^{mx}\varepsilon ,  \label{33000} \\
T_{11} &=&(\varepsilon +p)u_{1}u_{1}+pg_{11}=pa^{2}, \\
T_{22} &=&(\varepsilon +p)u_{2}u_{2}+pg_{22}=a^{2}e^{2mx}\left[ \varepsilon
\sigma +p(\sigma +k)\right] , \\
T_{33} &=&(\varepsilon +p)u_{3}u_{3}+pg_{33}=pa^{2}.
\end{eqnarray}

Now, it is convenient rewrite the field equations of the TEGR (3) as

\[
e_{a\lambda }e_{b\mu }\partial _{\nu }\left( ee_{d}{}^{\lambda }e_{c}{}^{\nu
}\Sigma ^{bdc}\right) -\left( e\eta _{ad}e_{c}{}^{\nu }\Sigma ^{bcd}T_{b\nu
\mu }-\frac{1}{4}e_{a\mu }e_{c}{}^{\gamma }e_{d}{}^{\nu }T_{b\gamma \nu
}\Sigma ^{bcd}\right) + 
\]

\begin{equation}
+\frac{1}{2}ee_{a\mu }\Lambda =\frac{1}{4k^{\prime }}ee_{a}{}^{\gamma
}T_{\gamma \mu },  \label{6500}
\end{equation}%
where:{\bf \ }%
\begin{eqnarray}
\Sigma ^{abc} &=&\frac{1}{4}\left( \eta ^{ad}e^{b\mu }e^{c\nu }T_{d\mu \nu
}+\eta ^{bd}e^{a\mu }e^{c\nu }T_{d\mu \nu }-\eta ^{cd}e^{a\mu }e^{b\nu
}T_{d\mu \nu }\right)  \nonumber \\
&&+\frac{1}{2}\left( \eta ^{ac}e^{b\nu }e^{d\mu }T_{d\mu \nu }-\eta
^{ab}e^{c\nu }e^{d\mu }T_{d\mu \nu }\right) .
\end{eqnarray}

The nonzero components of the torsion tensor $T_{a\mu \nu }$\ are given by%
\begin{eqnarray}
T_{(1)01} &=&-T_{(1)10}=T_{(3)03}=-T_{(3)30}=\dot{a},  \label{39000} \\
T_{(2)02} &=&-T_{(2)20}=\dot{a}\sqrt{k+\sigma }e^{mx},  \label{40000} \\
T_{(0)02} &=&-T_{(0)20}=\dot{a}\sqrt{\sigma }e^{mx},  \label{40500} \\
T_{(0)12} &=&-T_{(0)21}=m\dot{a}\sqrt{\sigma }e^{mx}, \\
T_{(2)12} &=&-T_{(2)21}=ma\sqrt{k+\sigma }e^{mx},  \label{41000}
\end{eqnarray}%
remembering that the torsion components are antisymmetrical under the
exchange of the two last indexes.

After tedious but straightforward calculations, we obtain the nonzero
components of the tensor $\Sigma ^{abc}$\ 
\begin{eqnarray}
\Sigma ^{(0)(1)(0)} &=&-\Sigma ^{(0)(0)(1)}=\frac{m}{2a},  \label{90500} \\
\Sigma ^{(3)(3)(1)} &=&-\Sigma ^{(3)(1)(3)}=\frac{m}{2a}, \\
\Sigma ^{(0)(1)(2)} &=&-\Sigma ^{(0)(2)(1)}=-\frac{m}{4a}\sqrt{\frac{\sigma 
}{k+\sigma }},  \label{90600} \\
\Sigma ^{(1)(2)(0)} &=&-\Sigma ^{(1)(0)(2)}=\frac{m}{4a}\sqrt{\frac{\sigma }{%
k+\sigma }}, \\
\Sigma ^{(2)(1)(0)} &=&-\Sigma ^{(2)(0)(1)}=-\frac{m}{4a}\sqrt{\frac{\sigma 
}{k+\sigma }},  \label{35000} \\
\Sigma ^{(1)(1)(0)} &=&-\Sigma ^{(1)(0)(1)}=-\frac{\dot{a}}{a},
\label{37000} \\
\Sigma ^{(2)(2)(0)} &=&-\Sigma ^{(2)(0)(2)}=-\frac{\dot{a}}{a},
\label{36000} \\
\Sigma ^{(3)(3)(0)} &=&-\Sigma ^{(3)(0)(3)}=-\frac{\dot{a}}{a},
\label{36900} \\
\Sigma ^{(0)(0)(2)} &=&-\Sigma ^{(0)(2)(0)}=-\frac{\dot{a}}{a}\sqrt{\frac{%
\sigma }{k+\sigma }},  \label{90700} \\
\Sigma ^{(1)(2)(1)} &=&-\Sigma ^{(1)(1)(2)}=-\frac{\dot{a}}{a}\sqrt{\frac{%
\sigma }{k+\sigma }},  \label{90800} \\
\Sigma ^{(3)(3)(2)} &=&-\Sigma ^{(3)(2)(3)}=\frac{\dot{a}}{a}\sqrt{\frac{%
\sigma }{k+\sigma }.}  \label{38000}
\end{eqnarray}

Next, we proceed to obtain the components of the field equations in the
model of the G\"{o}del-Obukhov universe. We mention here only the details
used to write the components $a=(0),\;\mu =0$ of field equations. We divide
the field equations into five parts. Thus we rewrite the field equations as%
\begin{eqnarray*}
Part\,1 &=&e_{(0)\lambda }e_{b0}\partial _{\nu }\left( ee_{c}{}^{\lambda
}e_{d}{}^{\nu }\Sigma ^{bcd}\right) \\
&=&e_{(0)0}e_{(0)0}\partial _{1}\left( ee_{\left( 0\right) }{}^{0}e_{\left(
1\right) }{}^{1}\Sigma ^{(0)(0)(1)}\right) + \\
&&e_{(0)0}e_{(0)(0)}\partial _{1}\left( ee_{\left( 1\right) }{}^{1}e_{\left(
2\right) }{}^{0}\Sigma ^{(0)(2)(1)}\right) + \\
&&e_{(0)2}e_{(0)(0)}\partial _{0}\left( ee_{\left( 2\right) }{}^{2}e_{\left(
0\right) }{}^{0}\Sigma ^{(0)(2)(0)}\right) + \\
&&e_{(0)2}e_{(0)(0)}\partial _{1}\left( ee_{\left( 2\right) }{}^{2}e_{\left(
1\right) }{}^{1}\Sigma ^{(0)(2)(1)}\right) \\
&=&-\frac{m^{2}ae^{mx}}{4}\left( \frac{2k+\sigma }{\sqrt{k+\sigma }}\right) -%
\frac{e^{mx}\sigma }{\sqrt{k+\sigma }}\left( a\dot{a}^{2}+a^{2}\ddot{a}%
\right) \\
&&\vdots \\
Part\,2 &=&-ee_{c}{}^{\nu }\eta _{(0)d}\Sigma ^{bcd}T_{b\nu 0} \\
&=&ee_{\left( 1\right) }{}^{1}\Sigma ^{(1)(1)(0)}T_{(1)10}+ee_{\left(
2\right) }{}^{2}\Sigma ^{(0)(2)(0)}T_{(0)20}+ee_{\left( 2\right)
}{}^{2}\Sigma ^{(2)(2)(0)}T_{(2)20}+ \\
&&ee_{\left( 3\right) }{}^{3}\Sigma ^{(3)(3)(0)}T_{(3)30} \\
&=&3\dot{a}^{2}a\sqrt{k+\sigma }e^{mx}-\frac{\dot{a}^{2}a\sigma e^{mx}}{%
\sqrt{k+\sigma }} \\
&&\vdots \\
Part\,3 &=&\frac{1}{4}ee_{a\mu }e_{c}{}^{\lambda }e_{d}{}^{\nu }T_{b\lambda
\nu }\Sigma ^{bcd} \\
&=&\frac{1}{4}ee_{a\mu }(2e_{\left( 0\right) }{}^{0}e_{\left( 1\right)
}{}^{1}T_{(1)01}\Sigma ^{(1)(0)(1)}+2e_{\left( 0\right) }{}^{0}e_{\left(
2\right) }{}^{2}T_{(2)02}\Sigma ^{(2)(0)(2)}+ \\
&&2e_{\left( 2\right) }{}^{0}e_{\left( 1\right) }{}^{1}T_{(1)10}\Sigma
^{(1)(1)(2)}+2e_{\left( 0\right) }{}^{0}e_{\left( 2\right)
}{}^{2}T_{(0)02}\Sigma ^{(0)(0)(2)}+ \\
&&2e_{\left( 0\right) }{}^{0}e_{\left( 2\right) }{}^{2}T_{(2)21}\Sigma
^{(0)(0)(2)}+2e_{\left( 3\right) }{}^{3}e_{\left( 2\right)
}{}^{0}T_{(3)30}\Sigma ^{(3)(3)(2)}+ \\
&&2e_{\left( 3\right) }{}^{3}e_{\left( 0\right) }{}^{0}T_{(3)30}\Sigma
^{(3)(3)(0)}) \\
&=&-\frac{k}{\sqrt{k+\sigma }}\frac{3\dot{a}^{2}ae^{mx}}{2}+\frac{%
m^{2}a\sigma e^{mx}}{8\sqrt{k+\sigma }} \\
&&\vdots \\
Part\,4 &=&\frac{1}{2}ee_{(0)0}\Lambda =-\frac{a^{3}\sqrt{k+\sigma }e^{mx}}{2%
}\Lambda \\
&&\vdots \\
Part\,5 &=&\frac{1}{4k^{\prime }}ee_{(0)}{}^{\nu }T_{\nu 0}=\frac{1}{%
4k^{\prime }}ee_{\left( 0\right) }{}^{0}T_{00}=\frac{a^{3}e^{mx}\varepsilon 
\sqrt{k+\sigma }}{4k^{\prime }}
\end{eqnarray*}%
Substituting Eqs. (23), (25), (26), (28), and (35)--(50) into the above
equation, and summing the parts, we obtain%
\begin{equation}
-\frac{2\sigma }{k+\sigma }\left( \frac{a\ddot{a}-\dot{a}^{2}}{a^{2}}\right)
+\frac{3\dot{a}^{2}}{a^{2}}\left( \frac{k}{k+\sigma }\right) -\frac{m^{2}k}{%
a^{2}\left( k+\sigma \right) }-\omega ^{2}-\Lambda =8\pi G\varepsilon ,
\label{45000}
\end{equation}%
where the dot indicates the temporal derivative and \ $\omega $\ is the
global angular velocity of the universe [31] given by 
\begin{equation}
\omega =\frac{m}{2a}\left[ \frac{\sigma }{(k+\sigma )}\right] ^{1/2}.
\end{equation}

Repeating the procedure for the components $a=(1),$ $\mu =0;$ $a=(1),$ $\mu
=1;$\ $a=(0)$, $\mu =2$; $a=(2)$, $\mu =0;$ and $a=(3),$ $\mu =3;$ we
obtain, respectively, the independent equations 
\begin{equation}
m\dot{a}\frac{\sigma }{\sqrt{k+\sigma }}=0,  \label{42000}
\end{equation}%
\begin{equation}
\frac{2a\ddot{a}+\dot{a}^{2}}{a^{2}}\left( \frac{k}{k+\sigma }\right)
-\omega ^{2}-\Lambda =-8\pi Gp,  \label{47000}
\end{equation}%
\begin{equation}
\frac{2a\ddot{a}+\dot{a}^{2}}{a^{2}}\left( \frac{k}{k+\sigma }\right) -\frac{%
m^{2}k}{a^{2}\left( k+\sigma \right) }-\omega ^{2}-\Lambda =8\pi
G\varepsilon ,  \label{49000}
\end{equation}

\begin{equation}
\frac{\left( \ddot{a}a-\dot{a}^{2}\right) }{a^{2}}\sqrt{\sigma }=0
\label{49500}
\end{equation}%
\begin{equation}
\frac{2a\ddot{a}+\dot{a}^{2}}{a^{2}}\left( \frac{k}{k+\sigma }\right) -\frac{%
m^{2}k}{a^{2}\left( k+\sigma \right) }-3\omega ^{2}-\Lambda =-8\pi Gp.
\label{50000}
\end{equation}

Equation (53) reveals the impossibility of the simultaneous existence of
rotation and expansion using a perfect fluid matter source for the G\"{o}%
del-Obukhov metric.

The field equations (53)--(57) represent the G\"{o}del-Obukhov universe in
the context of the TEGR. Fixing $m=0,$\ $\sigma =0$\ and $k=1,$\ they reduce
to the Friedmann equations for the flat universe%
\begin{eqnarray}
\frac{3\dot{a}^{2}}{a^{2}}-\Lambda &=&8\pi G\varepsilon ,\;\;\;\;\;\; 
\nonumber \\
\frac{2a\ddot{a}+\dot{a}^{2}}{a^{2}}-\Lambda &=&-8\pi Gp.
\end{eqnarray}

Moreover, Eqs. (53)--(57) are according to the works of Krechet \& Panov
[41], and Korothii \& Obukhov [42], using the Einstein equations and a
perfect fluid

\section{Total energy of the G\"{o}del-Obukhov model}

Let us now calculate the total energy of the G\"{o}del-Obukhov universe
using the equations shown in Section 2. By making $\lambda =0,$ and
substituting Eq. (5) in (12) and using 
\begin{equation}
\Sigma ^{a\lambda \nu }=\Sigma ^{abc}e_{b}{}^{\lambda }e_{c}{}^{\nu },
\label{eq413}
\end{equation}%
we have 
\begin{equation}
P^{a}=4k^{\prime }\int_{V}d^{3}x\partial _{i}\left( e\;\Sigma
^{abc}e_{b}{}^{0}e_{c}{}^{i}\right) .  \label{9000}
\end{equation}

As already observed, the temporal component represents the system energy.
Therefore the energy will be given by 
\begin{equation}
P^{(0)}=4k^{\prime }\int_{V}d^{3}x\partial _{i}\left( e\Sigma
^{(0)bc}e_{b}{}^{0}e_{c}{}^{i}\right) .
\end{equation}

Such a quantity can be written as 
\begin{equation}
P^{(0)}=4k^{\prime }\int_{V}d^{3}x\partial _{1}\left( ee_{\left( 0\right)
}{}^{0}e_{\left( 1\right) }{}^{1}\Sigma ^{(0)(0)(1)}+ee_{\left( 2\right)
}{}^{0}e_{\left( 1\right) }{}^{1}\Sigma ^{(0)(2)(1)}\right) .  \label{eq416}
\end{equation}%
By substituting Eqs. (25), (26), (40), and (42) in (62), we obtain 
\begin{equation}
P^{(0)}=-k^{\prime }am^{2}\frac{2k+\sigma }{\sqrt{k+\sigma }}%
\int_{V}d^{3}xe^{mx}.
\end{equation}%
Simplifying the above expression, it follows that the energy density is
given by 
\begin{equation}
{\mbox{\Large$\varepsilon$}}^{(0)}=-\frac{am^{2}}{16\pi G}\frac{2k+\sigma }{%
\sqrt{k+\sigma }}\,e^{mx}.  \label{51000}
\end{equation}%
We note that ${\mbox{\Large$\varepsilon$}}^{(0)}$ is negative and clearly
diverges when integrated in the entire space.

We can now compare our results with those obtained by Rybn\'{\i}ckov\'{a}
[11] and Dabrowski \& Garecki [10].

Rybn\'{\i}ckov\'{a} obtained the total energy density, $\omega ^{0}$, using
the Komar superpotential, as 
\[
\omega ^{0}=-\frac{a\sigma m^{2}e^{mx}}{16\pi \sqrt{k+\sigma }}. 
\]

Dabrowski and Garecki used the pseudotensor of Einstein for the stationary
metric of the acausal G\"{o}del model and arrived at the null result for the
total energy density [10]. They found a negative total energy density for
the causal G\"{o}del model. Sharif [43] calculated the total energy density
associated with a space-time homogeneous G\"{o}del-type metric by using
Einstein and Papapetrou energy-momentum complexes. Sharif%
\'{}%
s results are not in agreement with our result and he found that the two
definitions of energy-momentum complexes do not provide the same result for
this type of metric.

Our result and that obtained by Rybn\'{\i}ckov\'{a} present the same
dependence on $a(t)$ and $x$, but differ by a constant factor. Moreover, we
see that ${\mbox{\Large$\varepsilon$}}^{(0)}=0$, in the case $k=-1/2,$ $%
m=\sigma =1,$ corresponding to the energy density of the G\"{o}del universe
(rotation only). Then, our result is compatible with those of the works of
Rybn\'{\i}ckov\'{a} and Dabrowski \& Garecki for the acausal model. They
also found a nonzero total density energy for the causal G\"{o}del model.

In particular, fixing\ $m=\sigma =0$\ (without rotation) and $k=1$\ all
these the results are identical to zero%
\[
{\mbox{\Large$\varepsilon$}}^{(0)}=\omega ^{0}=0, 
\]%
recovering the result for the flat FLRW universe.

These results for the flat universe were also found by Rosen [3],
Cooperstock [4], Johri [5], and Garecki [6].

\section{The total momentum of the G\"{o}del-Obukhov model}

Let us now consider the calculation of the total three-momentum (matter plus
gravitational field) of the FLRW Universe. As seen in Section 2, it is noted
that the total three-momentum is given by space components $a=\{1\}$, $\{2\}$%
\ and $\{3\}$\ of Eq. (60) .

In order to obtain the space component $a=\{1\}$\ of the total momentum, we
can write the quantity $P^{(1)}$\ as

\begin{equation}
P^{(1)}=4k^{\prime }\int_{V}d^{3}x\partial _{1}(ee_{\left( 0\right)
}{}^{0}e_{\left( 1\right) }{}^{1}\Sigma ^{(1)(0)(1)}+ee_{\left( 2\right)
}{}^{0}e_{\left( 1\right) }{}^{1}\Sigma ^{(1)(2)(1)}).
\end{equation}

By substituting (25), (26), (45), and (49) in the previous equation, we
obtain 
\begin{equation}
P^{(1)}=\frac{4k^{\prime }\dot{a}amk}{\sqrt{k+\sigma }}\int_{V}d^{3}xe^{mx}.
\label{52000}
\end{equation}

Thus, the spatial momentum density \ $\wp ^{(1)}$ it is given by 
\begin{equation}
\wp ^{(1)}=\frac{a\dot{a}mk}{4\pi G\sqrt{k+\sigma }}\,e^{mx}.  \label{53000}
\end{equation}

For the absolutely analogous calculation, we have%
\begin{equation}
\wp ^{(2)}=\frac{m^{2}a\sqrt{\sigma }}{16\pi G}\,e^{mx},  \label{54000}
\end{equation}%
\begin{equation}
\wp ^{(3)}=0.  \label{55000}
\end{equation}

We can observe that Eq. (67) is valid for any fluid matter. However, it
shows that considering a perfect fluid, for a purely rotational or expansion
universe, the spatial momentum density $\wp ^{(1)}$ is zero. Equation (68)
shows that the spatial momentum density $\wp ^{(2)}$ is zero only for an
expansion universe. We can note again that $\wp ^{(1)}$ and $\wp ^{(2)}$
diverge when integrated in the entire space.

\section{The Gravitational angular momentum of the G\"{o}del-Obukhov model}

Let us verify the expression of gravitational angular momentum (18). By
making use of (19) and (17), we can write (18) in the form 
\begin{equation}
L^{ab}=-4k^{\prime }\int_{V}d^{3}xe\left( \Sigma ^{a0b}-\Sigma ^{b0a}\right)
.  \label{eq58}
\end{equation}%
By making use of $\Sigma ^{a0b}=e_{c}{}^{0}\Sigma ^{acb}$\ and using the
determinant of tetrads, Eq.(26), and the components of the tensor $\Sigma
^{abc}$\ , (40)--(50), we find that the components of the three-angular
momentum densities ${\cal L}^{(1)(3)}$ and ${\cal L}^{(2)(3)}$ vanish. The
unique nonzero gravitational three-angular momentum density component is
given by

\begin{equation}
{\cal L}^{(1)(2)}=-{\cal L}^{(2)(1)}=\frac{ma^{2}\sqrt{\sigma }}{8\pi G}%
e^{mx}.  \label{81500}
\end{equation}%
The above result demonstrates that there is only a direction to the
gravitational angular momentum, as expected. The component ${\cal L}%
^{(1)(2)} $ found here diverges when integrated in the entire space.

The other components ${\cal L}^{(0)(i)}$ are all null except for

\begin{equation}
{\cal L}^{(0)(1)}=-{\cal L}^{(1)(0)}=\frac{ma^{2}\sqrt{k+\sigma }}{8\pi G}%
e^{mx},  \label{81600}
\end{equation}%
which, although it represents the component of the gravitational center of
mass moment, does not possess physical meaning [44].

Our result Eq.(71) when fixing $k=-1/2,$ $m=\sigma =1$ is not in agreement
with those obtained by Dabrowski \& Garecki which used the gravitational
angular momentum pseudotensor of Bergmann-Thomson to calculate the
gravitational angular momentum density [10]. Indeed as they used the
pseudotensor to calculate the gravitational angular momentum, their results
are not coordinate invariant.

\section{Conclusions}

In this work, we show explicitly the equivalence among equations obtained
with the TEGR and those obtained by the GR for the G\"{o}del-Obukhov metric
and also calculated the total energy-momentum and gravitational angular
momentum densities with the use of tensorial expressions of the TERG,
irrespective of the equations of state of the cosmic fluid metric. In the
case in which the rotation parameters equals to zero, we recovered the flat
FLRW universe.

Our result for the energy density presents the same dependence on $a(t)$ and 
$x$ as that found by Rybn\'{\i}ckov\'{a}. Both results diverge when
integrated in the entire space. Fixing $k=-1/2,$ $m=\sigma =1$ (parameters
of the stationary G\"{o}del metric), the energy density vanishes in
accordance with the results obtained by Dabrowski \& Garecki.

By analyzing the equations for the total three-momentum (matter plus
gravitational field) we conclude that all these components vanish
simultaneously in the universe without rotation.

As we used tetrads fields adapted to static observers in space-time, we
found a unique nonzero component of the gravitational angular momentum
density with physical meaning, ${\cal L}^{(1)(2)}$, which reflects the
preferred direction (z-axis) related to the cosmic rotation.

We finally conclude from this work that the TEGR obtained equivalent results
to the GR with the great advantage of addressing covariantly the definitions
of quantities as energy-momentum and angular momentum tensors of the
gravitational field.

In order to find a solution that simultaneously permits rotation and
expansion, we will consider others types of fluids as the Chaplygin gas.
Efforts in this respect will be carried out.

\bigskip \noindent Acknowledgements

\noindent One of us (A. C. S.) would like to thank the Brazilian agency\
CAPES by financial support.

\end{document}